\documentclass[twocolumn,showpacs,preprintnumbers,amsmath,amssymb]{revtex4}
\usepackage{dcolumn}
\usepackage{bm}

\usepackage{graphicx}

\newcommand{\rz}{\rangle}
\newcommand{\lz}{\langle}

\newcommand{\ha}{ {a}}
\newcommand{\had}{ {a}^{\dag}}

\newcommand{\be}{\begin{equation}}
\newcommand{\ee}{\end{equation}}
\newcommand{\bea}{\begin{eqnarray}}
\newcommand{\eea}{\end{eqnarray}}

\newcommand{\FIG}{\begin{figure}[h]\begin{center}}
\newcommand{\FIGe}{\end{center}\end{figure}}
\newcommand{\DN}{\delta N_{\rm G}}
\newcommand{\DE}{\delta \epsilon}

\newcommand{\qp}{\textrm{qp}}
\newcommand{\Gone}{\Gamma_{1,n\rightarrow\pm,n+1}^{\qp}}
\newcommand{\Gpm}{\Gamma_{\pm,n\rightarrow 1,n}^{\qp}}

\newcommand{\Eone}{\Delta E_{1,n\rightarrow\pm,n+1}}

\newcommand{\Eonem}{\Delta E_{1,n\rightarrow -,n+1}}
\newcommand{\Ng}{N_{\textrm{G}}}

\begin{document}

\title{Photon-Number Squeezing in Circuit Quantum Electrodynamics}

\author{ M.~Marthaler$^1$, Gerd Sch\"on$^1$, and Alexander Shnirman$^2$}

\affiliation{$^1$Institut f\"ur Theoretische Festk\"orperphysik
  and DFG-Center for Functional Nanostructures (CFN), Universit\"at Karlsruhe, D-76128 Karlsruhe, Germany\\
  $^2$Institut f\"ur Theoretische Physik, Universit\"at Innsbruck, A-6020 Innsbruck, Austria}
%
\date{\today}

\begin{abstract}
A superconducting single-electron transistor (SSET) coupled to an
anharmonic oscillator, e.g., a Josephson junction-$L$-$C$ circuit,
can drive the latter to a nonequilibrium photon number state. By
biasing the SSET in a regime where the current is carried by a
combination of inelastic quasiparticle tunneling and coherent
Cooper-pair tunneling (Josephson quasiparticle cycle), cooling of
the oscillator as well as a laser like enhancement of the photon
number can be achieved. Here we show, that the cut-off in the
quasiparticle tunneling rate due to the superconducting gap, in
combination with the anharmonicity of the oscillator, may create
strongly squeezed photon number distributions.~For low dissipation
in the oscillator nearly pure Fock states can be produced.
\end{abstract}


\maketitle

 In driven oscillator systems, depending on the type of driving,
 different nonequilibrium photon (or phonon) populations can be produced, and indeed for many applications the
 production of specific photon number distributions is crucial.
 Quantum cryptography and linear optical quantum computation
 require a reliable supply of single photons~\cite{Gisin2002,Knill2001},
 whereas for quantum measurements it
 may be of advantage to use strongly squeezed photon
 distributions~\cite{Giovannetti2004}. A well-known optical system where
 strongly squeezed states  can be created
 is the micromaser~\cite{Varcoe2000}. In this article we describe how highly
 squeezed photon (phonon) number distributions can be produced in suitable
 superconducting quantum circuits.

Recent experiments with such circuits, with superconducting qubits coupled to
 electromagnetic resonators (``circuit QED"), confirmed
 various concepts developed in the field of cavity QED
 but also provided important extensions, e.g., to the regime of
 strong coupling~\cite{Wallraff2004,Delft2004,Schuster2005,Schuster2007,Blais2004}.
 A Josephson qubit, ac-driven to perform Rabi
 oscillations in resonance with an oscillator, depending
 on the detuning either cools the oscillator or produces a
 laser-like enhancement of the photon numbers~\cite{Jena2003,Hauss2007}.
 Similarly, a superconducting SET (SSET) biased at
 the Josephson quasiparticle (JQP) cycle can be used to cool or
 drive an oscillator \cite{Astafiev2007,Blencowe2005,Clerk2005,Rodriguess2005}.
 Squeezing of the photon number distribution has been predicted, but
 as described in the literature it is only a weak effect
 \cite{Rodrigues2007,Hauss2007}. Here we show that by exploiting
 the gap structure of the quasiparticle tunneling rate in
 combination with an anharmonicity of the oscillator, strongly squeezed
photon number states, close to a pure Fock state
 can be produced (see fig.~\ref{fig:squeezed}).

\FIG
 \includegraphics[width=2.5 in]{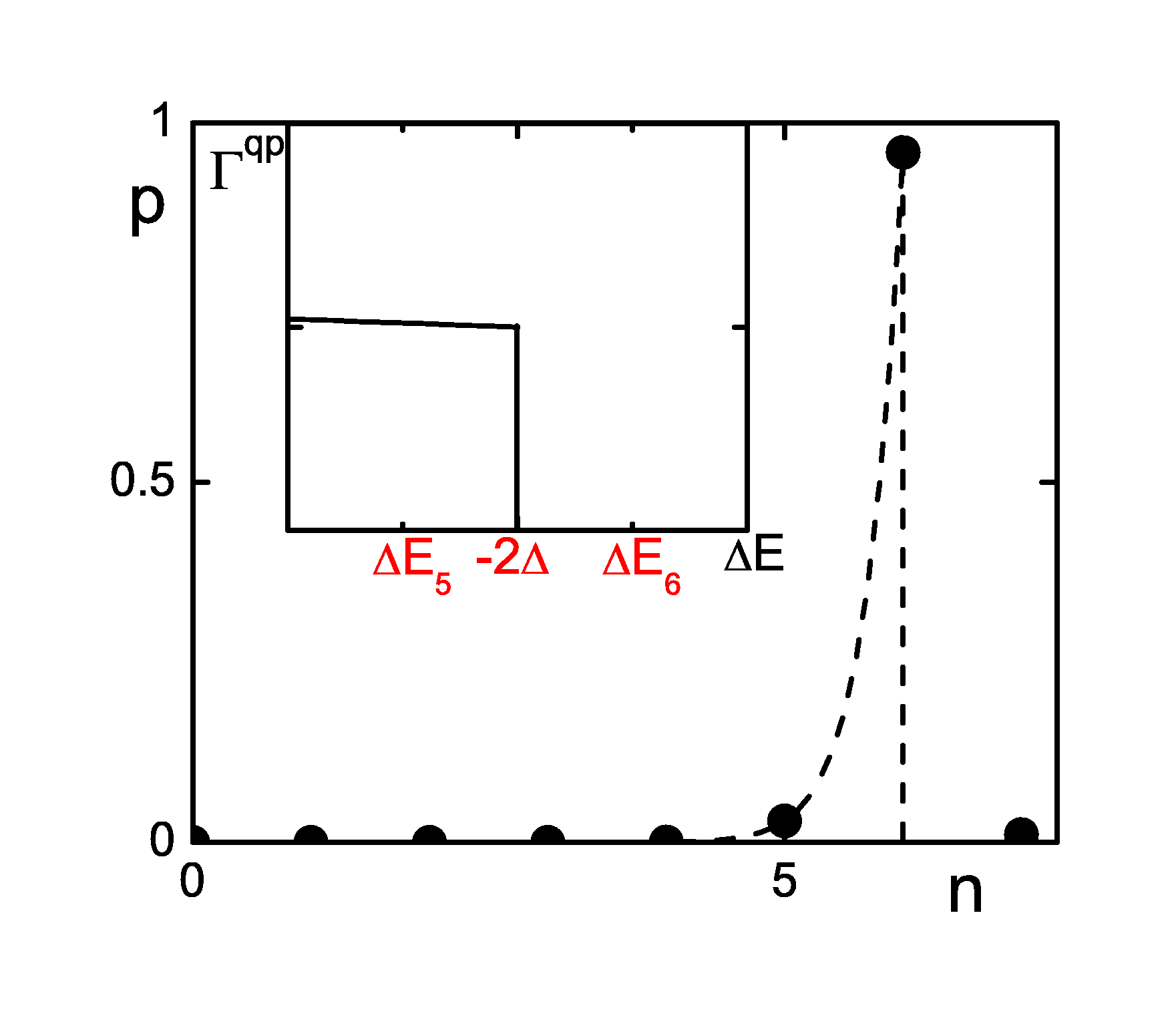}
 \vspace{-5mm}
\caption{A strongly squeezed distribution
 $p(n)$ of photon number states in a Josephson junction-$L$-$C$
 oscillator driven by a SSET obtained for the parameters
 $\delta N_{\rm G}=-0.1$, $eV=5.6$, $\Delta=2.2$,
 $g=0.01$, $\omega=0.4$, $\Omega=0.001$, $E_{\rm J}=0.1$
 (all energies in units of $E_C$), and $\kappa/\gamma=0.004$.
 The inset shows the energy dependence of the quasiparticle
 transition rate and energy differences for two transitions,
 $\Delta E_n=E_{-,n+1}-E_{1,n}-eV$. Due to the anharmonicity of
 the oscillator they lie above and below the threshold.}\label{fig:squeezed}
 \FIGe

 \begin{figure}[t]
 \begin{center}
 \includegraphics[width=2.5 in]{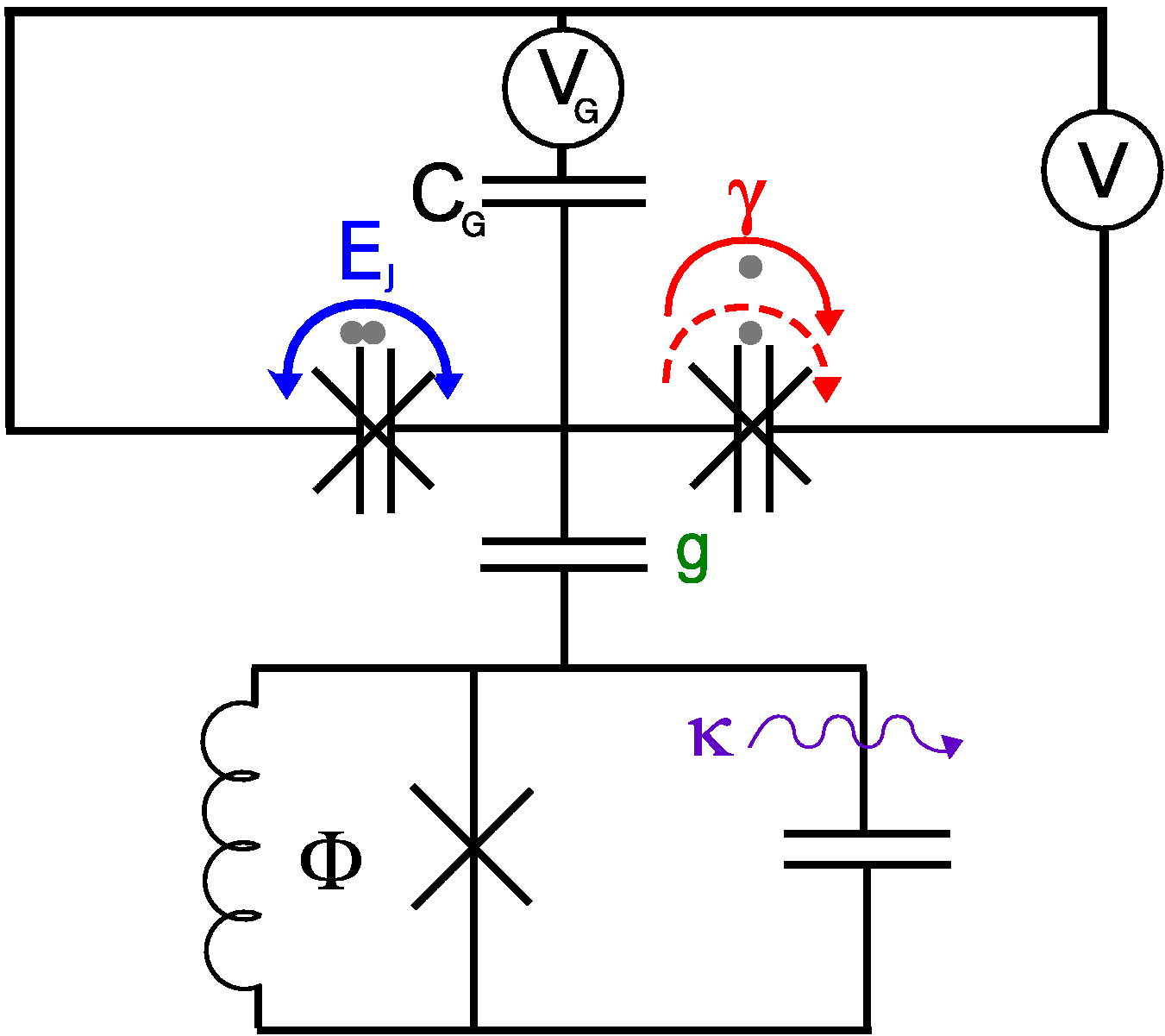}
 \caption{ A SSET with Josephson coupling $E_{\rm J}$ and quasiparticle tunneling
 rate proportional to $\gamma$, coupled with strength $g$ to an anharmonic
 oscillator.
 The oscillator's dissipation is characterized by the parameter $\kappa$.
 To get the right sign for the anharmonicity we use the flux $\Phi$. The current
 is determined by the transport voltage V and the gate voltage $V_G$ which is coupled
 to the island by the capacitance $C_G$.}\label{fig:system}
 \end{center}
 \end{figure}

 \emph{The system} to be studied consists of a SSET coupled to an anharmonic
 superconducting or nanomechanical oscillator (see
 fig.~\ref{fig:system}). The SSET is formed by a superconducting island
 coupled via low-capacitance tunnel junctions to two superconducting leads.
 A gate voltage $V_{\rm G}$ shifts the
 electrostatic energy of the island and controls,
 together with the transport voltage $V$, the current through the device.
 The Josephson coupling $E_{\rm J}$ of the junctions should be small compared
 to the charging energy scale, $E_C= e^2/2C$ ($C$ is the total
 capacitance of the island),
 and the superconducting gap $\Delta$. It leads
 to coherent Cooper pair tunneling, and even in the considered limit, has pronounced consequences when two
 charge states differing by one Cooper-pair are nearly degenerate.
 In addition, quasiparticles can tunnel incoherently (with rate
 $\propto \gamma$)  when
 the energy difference between initial and final states
 is sufficient to create the quasiparticle excitation, i.e., when it
 exceeds twice the superconducting energy gap
 (assumed equal for electrodes and island), $|\Delta E |\ge2\Delta$.
 At the threshold tunneling sets in with a sharp step.

 The SSET is tuned to the regime of the JQP cycle~\cite{Fulton1989,Maassen1991},
 where the current is carried by a combination of
 Cooper pair transfer through one junction onto the island
 followed by two consecutive
 quasiparticle tunneling processes through the other junction.
 The energy for this process is provided by the voltage source.
 In oder to enhance the effect we consider in the following an asymmetric SSET,
 similar to those studied in Ref.~\cite{Astafiev2007}, with one junction
 (where the Cooper pair is transferred during the JQP cycle)
 having a much stronger Josephson coupling than the other one.

 The charge on the SSET is coupled capacitively, with strength characterized by $g$, to an anharmonic oscillator. This oscillator can be realized, e.g., by
 a circuit combining a capacitor, an inductor,
 and a non-linear element such as a Josephson junction,
 as shown in fig.~\ref{fig:system}.
 To get strong squeezing we need a
 positive quartic term in the potential energy,
 which can be achieved by shifting the minima of
 the inductive energy and Josephson coupling relative to each other by
 biasing the circuit with a half-integer number
 of flux quanta. Also nanomechanical
 oscillators must have an anharmonicity with positive sign
(as studied, e.g., in Ref.~\cite{Almog2007}) for the
 phonon number squeezing mechanism described below to work.

 The coherent dynamics of the coupled SSET
 and oscillator is described by the Hamiltonian
 \bea
   H_0 &=& E_C( {N}-\Ng)^2-E_{\rm J}\cos\left(\phi_{\rm L}\right)\nonumber\\
 &+&g\left( {N}-1\right)\left(\had+\ha\right)+\omega\had\ha+\Omega\left(\had\ha\right)^2\, .
 \eea
 The charging energy depends on the number of charges on the island,
 $  N=  N_{\rm L}-  N_{\rm R}$, where $  N_{\rm L/R}$
 are the numbers of charges which have tunnelled across the
 left/right junction and the gate charge $N_G=C_G V_G$.
 Cooper pair tunneling, assumed to dominate across the left junction,
 depends on the phase $\phi_{\rm L}$, which is conjugate
 to $  N_{\rm L}$, $[  N_{\rm L},e^{i\phi_{\rm L}}]=2e^{i\phi_{\rm L}}$.
 The oscillator's eigenfrequency $\omega$ should be of the same order
 as the Josephson coupling $E_{\rm J}$. The anharmonicity $\Omega$ is weak,
 $\Omega \lz n^2 \rz \ll \omega \lz n \rz$,
 where $\lz n \rz$ is the average number of photons. The oscillator
 couples to the SSET with strength $g$. Without loss of generality we choose
 the oscillator to be at its equilibrium position for the island
 charge $N=1$.

 In addition to the coherent dynamics, governed by $  H_0$,
 the state of the system evolves due to incoherent quasiparticle
 tunneling in the SSET and due to dissipative processes in the
 oscillator. These effects will be described in the frame of a Liouville equation.

 \begin{figure}[b]
 \begin{center}
 \includegraphics[width=3.2 in]{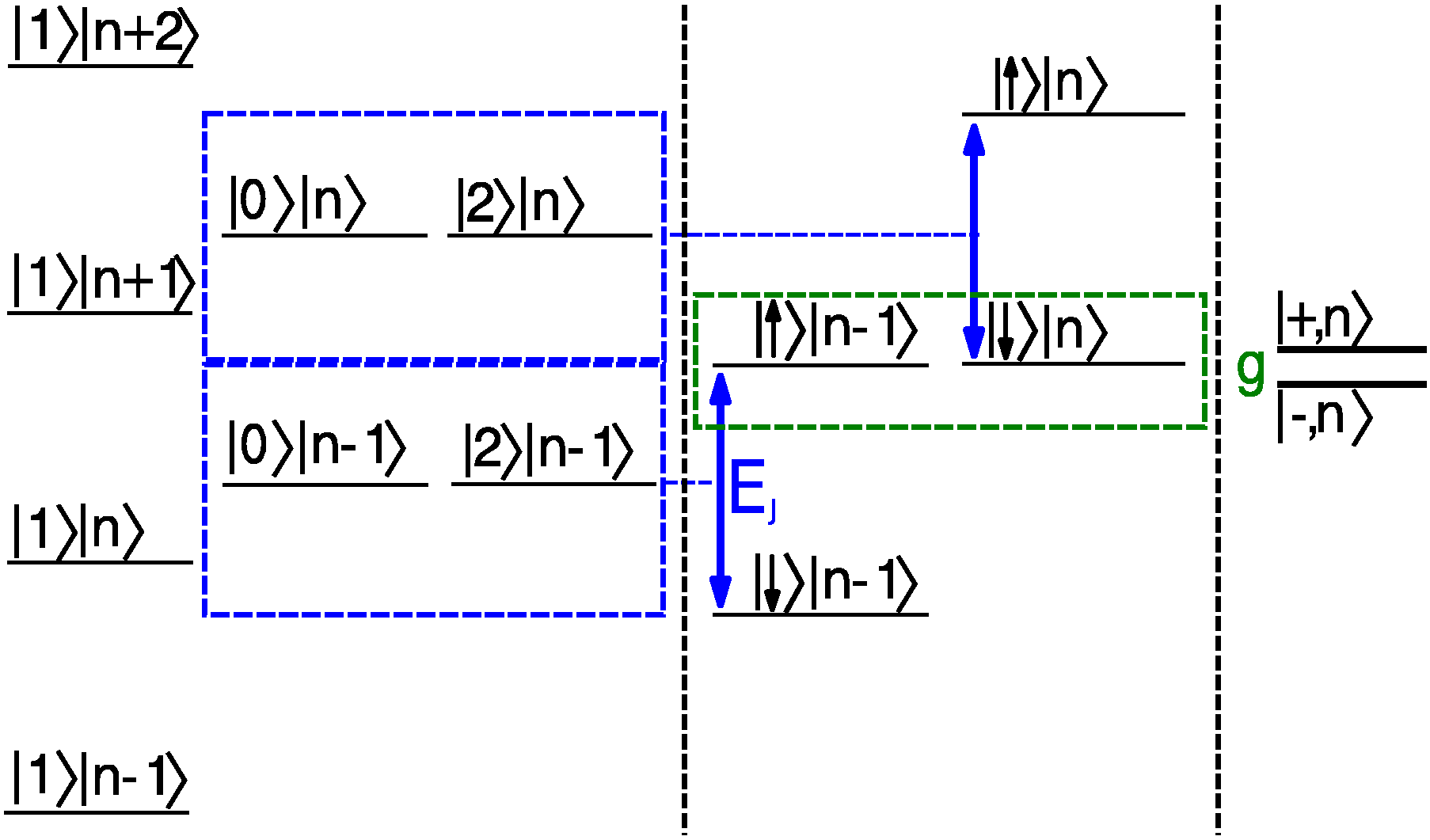}
 \caption{The charge states $|N=0\rangle$ and $|N=2\rangle$
 (here shown exactly at degeneracy) are
 coupled by the Josephson coupling $E_{\rm J}$ and form
 the basis states $|\uparrow\rangle$ and $|\downarrow\rangle$.
 Tuned to a resonance with the oscillator they form the
 eigenstates $|\pm,n \rangle$ of the coherent part of the Hamiltonian.
 Also the odd states $|1\rz |n\rz$ are shown. Their
 energy is shifted by an amount of the order of $E_C$ relative
 to the qubit states, but to fit into the plot this shift is not drawn to scale.
 }\label{fig:levels}
 \end{center}
 \end{figure}

 \begin{figure}[t]
 \begin{center}
 \includegraphics[width=3.3 in]{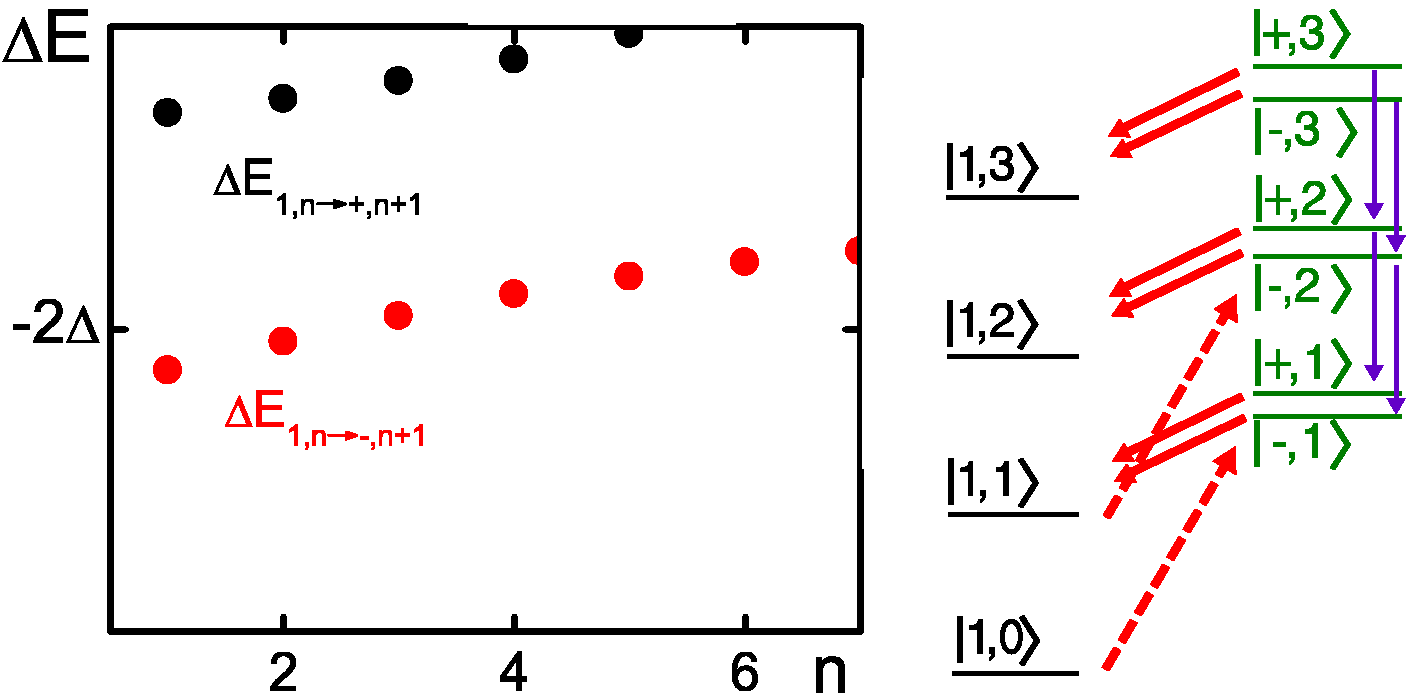}
 \caption{The energy differences $\Eone$
 as a function of photon number $n$ for the same parameters $\delta \omega$,
 $\Omega$ and $\bar{g}$ as used in fig.~\ref{fig:squeezed}.
 Also shown is a cycle of transitions,
 which increases the number of photons in the oscillator.
 The dashed arrows correspond to the rate $\Gone$ and the
 solid arrows correspond to  the rate $\Gpm$.
 The vertical transitions are due to
 the dissipation decreasing the number of photons.}
 \label{fig:rates}
 \end{center}
 \end{figure}

 We tune the gate charge close to a symmetry point,
 $N_{\rm G} \approx -1$, such that only the charge states $N=0,1,2$ are
 of importance~\cite{Makhlin2001}. The two even states, $N=0,2$,
 are Josephson coupled to form the basis states
 \bea
 |\uparrow\rz &=& \cos\frac{\chi}{2}|N=0\rz-\sin\frac{\chi}{2}|N=2\rz\nonumber\, ,\\
 |\downarrow\rz &=& \sin\frac{\chi}{2}|N=0\rz+\cos\frac{\chi}{2}|N=2\rz\, ,
 \eea
 with energies
 \bea
 \epsilon_{\uparrow/\downarrow}=(1+\delta N_{\rm G}^2) E_C
 \pm\frac{1}{2}\sqrt{E_{\rm J}^2 + 16\delta N_{\rm G}^2 E_C^2}\, .
 \eea
 Here  $\tan\chi=E_{\rm J}/4\delta N_{\rm G} E_C$,
 and $\delta N_{\rm G}=N_{\rm G}-1$ is the deviation  from the symmetry point.
 In addition we consider the odd state $|N=1\rz$
 with energy $\epsilon_1=E_C \delta N_{\rm G}^2$.

 We assume that the system is operated in the regime of vacuum Rabi oscillations,
 where the energy difference of the SSET states
 $\DE=\epsilon_{\uparrow}-\epsilon_{\downarrow}$
 is close to the oscillator frequency $\omega$, and the SSET and
 photon number states $|n\rz$ get strongly entangled.
 For weak detuning $\delta \omega=\omega-\delta \epsilon$
 we can approximate the eigenstates
 of $  H_0$ by
 \bea
 |+,n\rz &=&  \sin\frac{\eta}{2}
 |\uparrow\rz|n-1\rz+\cos\frac{\eta}{2}|\downarrow\rz|n\rz\,,\nonumber\\
 |-,n\rz &=&
 \cos\frac{\eta}{2} |\uparrow\rz|n-1\rz-\sin\frac{\eta}{2}|\downarrow\rz|n\rz\,.\label{eq:pmnStatesofH0}
 \eea
 The rotation angle, $\tan \eta=2\bar{g}\sqrt{n}/\delta E(n)$, depends
 on the effective detuning, $\delta E(n)=\delta\omega+\Omega(2n-1)$,
 and the effective coupling,
 $\bar{g}=g\lz \uparrow|( {N}-1)|\downarrow\rz$.
 In addition, the states $|N=1\rz|n\rz$
 with a single excess charge on the island are eigenstates of the system.
 The energies of the relevant states are then given by
 \bea\label{eq:endiff}
 E_{\pm,n} &=& \epsilon_{\uparrow}+E_{\rm{osc}}(n-1) + \frac{1}{2} \delta E(n)\nonumber\\
 & & \pm \frac{1}{2}\sqrt{4\bar{g}^2 n+ \delta E(n)^2}\,,\nonumber\\
 E_{1,n} &=& \epsilon_{1}+E_{\rm{osc}}(n)\,,
 \eea
 where $E_{\textrm{osc}}(n)=\omega n+\Omega n^2$
 is the energy of the anharmonic oscillator.

 \begin{figure*}[t]
 \begin{center}
 \includegraphics[angle=-90,width=7 in]{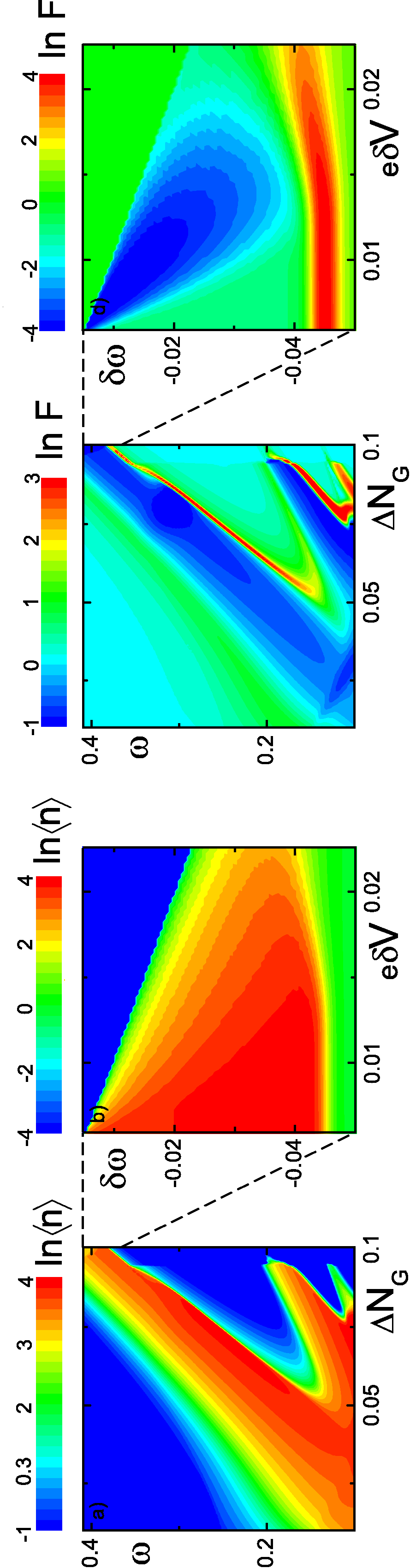}
 \caption{(a) and (c): Average photon number $\lz n \rz$ and Fano factor $F$
 as a function of the oscillator frequency $\omega$ and the gate charge
 $\Delta N_0$, for $eV=5.62$, $g=0.01$, $\Omega=0.0005$, $E_{\rm J}=0.18$,
 $\kappa/\gamma=0.02$ (all energies in units of $E_C$).
 (b) and (d): Average photon number
 $\lz n \rz$ and Fano factor $F$ as a function of the frequency detuning
 $\delta \omega$ and $e\delta V$ for $\Delta N_0=-0.1$.}\label{fig:Results}
 \end{center}
 \end{figure*}

 \emph{The Liouville equation} for the reduced density
 matrix of the composite system accounts for dissipative processes.
 Assuming the transition rates to be weak compared to the energy spacings
 we can use the rotating wave approximation and neglect
 off-diagonal elements. The probability of
 the system to be in the state $|i\rz$
 is then determined by the simple master equation
 \bea\label{eq:balance1}
 \dot{\rho}_i=\sum_j\left(\Gamma_{j\rightarrow i}\rho_j-\Gamma_{i\rightarrow j}\rho_i\right)\,.
 \eea
 The transition rates
 $\Gamma_{i\rightarrow j}=\Gamma_{i\rightarrow j}^{\textrm{qp}}+\Gamma_{i\rightarrow j}^\kappa$
 account for quasiparticle tunneling and for transitions caused
 by the dissipation of the oscillator. Since both have been
 described earlier, e.g.\ in Refs.~\cite{Rodrigues2007,Choi2003},
 it is sufficient to quote the results.

 The quasi-particle tunneling rate, assumed to dominate in the right junction, is given by
 \bea\label{eq:Gqp}
  \Gamma_{i\rightarrow j}^{\textrm{qp}}
 = |\lz
 j|e^{i\phi_{\rm R}/2}|i \rz|^2 I_{\qp}(\Delta E_{i\rightarrow j})\, .
 \eea
 The operator $e^{i\phi_{\rm R}/2}$ transfers a single
 charge from the island to the right lead. (For the bias and low
 temperatures considered single charges tunnel only in this direction.)
 The relevant energy difference
 $\Delta E_{i\rightarrow j}=E_j-E_i-eV$  also includes the work provided by the voltage source.
 The strength and energy dependence
 of the rate are related to the quasiparticle tunneling current,
 $I_{\qp}(E)= \gamma \int dE' N(E')N(E'-E)f(E')[1- f(E'-E)]$,
 which depends on  Fermi functions and the densities of states,
 $N\left(E\right)=\Theta\left(|E|-\Delta\right)|E|/\sqrt{E^2-\Delta^2}$.
 At low temperature we need to consider only inelastic
 transitions where energy is gained, i.e., $\Delta E_{i\rightarrow j} <0$.
 Due to the gap in the DOS
 the rate vanishes for $\Delta E_{i\rightarrow j}>-2\Delta$,
 but jumps to a finite value if
 $\Delta E_{i\rightarrow j}\le -2\Delta$.
 For an anharmonic oscillator this energy difference
 depends on the photon number $n$. Hence at a
 certain value of $n$ the threshold may be passed,
 beyond which the transitions vanish abruptly (as shown in Fig.~1.)
 This cut-off is essential for the creation of
 a strongly squeezed photon distribution.

 The dissipation in the oscillator introduces further transitions,
 \bea
 \Gamma_{i\rightarrow j}^\kappa
 &=&\frac{\kappa|\lz j|\ha|i \rz|^2}{|1-\exp([E_j-E_i]/kT)|}\, .
 \eea
 The parameter $\kappa$ summarizes all details of the
 oscillator's dissipation at the energy scale $E_j-E_i$.
 Most important are transitions with energy differences of the
 order of $\omega$. We consider low temperatures, where only
 relaxation processes occur.

 If the transport voltage $V$ is large compared to the gap,
 the number of photons in the oscillator increases for positive $\DN$,
 until a balance between
 driving and dissipation - which is proportional to $n$ - is reached.
 The distribution of photon numbers is peaked around the average
 value $n_{\textrm{av}}\propto \gamma/\kappa$,
 and the Fano factor $F=(\lz n^2\rz-{\lz n\rz}^2)/\lz n \rz$
 is slightly smaller than one. In this situation one observes some
 photon number squeezing~\cite{Rodrigues2007}, however it is weak and
 easily destroyed by temperature.
 For negative detuning $\DN$ the SSET, similar as an ac-driven qubit~\cite{Hauss2007}, can serve to cool the oscillator.

As the voltage is decreased certain rates can be pushed beyond the threshold
 at $\Delta E_{i\rightarrow j}=-2\Delta$. The rates with the largest energy
 differences $\Eone$ will be cut-off first. These are given by
 \bea
 \Eone &=&\epsilon_{\uparrow}-\epsilon_{1}-eV+\frac{1}{2}\delta E(n+1)\\
 & &\pm\frac{1}{2}\sqrt{4\bar{g}^2(n+1)+[\delta E(n+1)]^2}\nonumber\, .
 \eea
 The upper
 branch will be cut-off first.
 The most interesting case is shown in fig.~\ref{fig:rates},
 where the cycle stops completely beyond a certain number of photons.
 We will now describe how to tune the system to reach this situation.

 \emph{Strong squeezing} requires that
 the energy difference $\Eonem$ increases with increasing photon number $n$
 (as is the case in fig.~\ref{fig:rates}).
 The condition
 $\frac{\partial}{\partial n} (\Eonem)>0$ requires
 \bea
 \Omega>\frac{1}{2}\left(\delta\omega+\sqrt{2\bar{g}^2+\delta\omega^2}\right)
 \label{eq:condition1}\,.
 \eea
 This condition is independent of $n$, i.e., the
 energy difference either decreases or increases with $n$ for all states.
 We can see from eq.~(\ref{eq:condition1}) that for a negative
 detuning,
 $\delta\omega <0$, an anharmonicity smaller than $\bar{g}$ is sufficient.

 Squeezing can be observed below
 a certain threshold for the transport voltage.
 For large $n$ there is a limiting value for $\Eonem$ which has to be larger
 than $-2\Delta$; otherwise there is no point where the rate gets cut-off.
 From this we get a condition for the voltage
 \bea \label{eq:condition2}
 \frac{1}{2}\left(\delta E(1)-\sqrt{4\bar{g}^2+[\delta E(1)]^2}\right)<e\delta V < -\bar{g}^2/2\Omega\, ,
 \eea
 with $e\delta V=eV-2\Delta+\epsilon_1-\epsilon_{\uparrow}$. The right hand inequality
 guarantees that there is a cut-off, and the left hand side guarantees that the
 cycle does not stop already at zero photons.
 If the conditions given by Eqs.~(\ref{eq:condition1}) and
 (\ref{eq:condition2}) are met, the rates are cut-off at
 \bea
 n_{\textrm{cut}}
 =\frac{e\delta V\left(e\delta V-\delta\omega+\Omega \right)}{\bar{g}^2+2e\delta V\Omega}\, .
 \eea
 In order to have significant effects of the cut-off,
 we also have to require that
 ${n_{\textrm{av}} > n_{\textrm{cut}}}$.

 We can optimize the system by choosing
 a negative detuning, $\delta \omega<0$,
precisely in a way
 that the system is in resonance at
 the cut-off, $\delta E(n_{\textrm{cut}})=0$.
 This means
 \bea\label{eq:dwoptimal}
 \delta\omega=\Omega[1-2(e\delta V)^2/\bar{g}^2]\, .
 \eea
 In this case we get $n_{\textrm{cut}}=(e\delta V)^2/\bar{g}^2$.

 We solved for the stationary distribution of eq.~(\ref{eq:balance1}) in the product base of the charge states $N=0,1,2$ and many Fock states with $n \le 200$ sufficient to guarantee convergence.
 In fig.~\ref{fig:Results} (a) and (c) we plot the average photon number and the
 Fano factor for a fixed transport voltage.
 The maxima in the photon number and the minima in the Fano factor
 correspond to $\omega=\DE$ or to higher order resonances.
 The region where the conditions
 (\ref{eq:condition1}) and (\ref{eq:condition2}) are fulfilled lies in
 the center of fig \ref{fig:Results} (d).
 Here we find strong squeezing.

 In fig.~\ref{fig:squeezed} we show the probability distribution
 for the oscillator states $p(n)=Tr(\rho|n\rz\lz n|)$
 for parameters which meet eqs.~(\ref{eq:condition1}), (\ref{eq:condition2})
and (\ref{eq:dwoptimal}).
 For these parameters the Fano factor is particularly small, $F\approx 0.01$,
 and one can clearly see the effect
 of the cut-off. The rates are cut at $n=7$.
 Therefore we have a sharp maximum in $p$ at $n=6$ and then
 a sudden drop. The probabilities above $n=6$ are not exactly zero,
 because the numerically calculated quantum states
 allow for more none-zero matrix elements in the transition rates (\ref{eq:Gqp})
 than the approximations given by
 eqs.~(\ref{eq:pmnStatesofH0}).
 However, these additional rates are much
 smaller than $\Gone$ and $\Gpm$,
 as can be seen by the significant
 drop in $p$ for $n>6$.

 \emph{Conclusion.} A strongly squeezed photon number
 distribution can be produced in a solid state
 anharmonic oscillator coupled to a SSET.
 It requires an oscillator with frequency in the GHz range, a
 positive quartic term, and sufficiently low dissipation,  such that the inequalities
$\omega>g>\gamma\gg\kappa$ are satisfied. For the example
presented in fig.~\ref{fig:Results} a Q-factor of the order of
$10^4$ is sufficient. Nanomechanical oscillators, with $Q$-Factors
of the order
 of $10^{5}$ have been produced~\cite{Naik2006}, but
 reaching the GHz range is difficult. Circuits formed of a
 Josephson junction and $L$-$C$ elements can satisfy the requirements
 and have the advantage of tunable anharmonicity and frequency,
 which in turn allows selecting the squeezed photon number state.
 Furthermore, these circuits can be coupled to superconducting
 qubits, which have been demonstrated to allow measuring the
 oscillator state as a state-dependent frequency
 shift~\cite{Schuster2005}.

 \emph{Acknowledgment:} We thank V.~Brosco, M.~P.~Blencowe,
O.~Astafiev, and Y.~Nakamura for stimulating discussion.
The work is part of the EU IST Project EuroSQIP.


\begin{thebibliography}{48}


 \bibitem{Gisin2002}
 N.~Gisin, G.~Ribordy, W.~Tittel, and H.~Zbinden,
 Rev.~Mod.~Phys.~{\bf 74}, 145 (2002).


 \bibitem{Knill2001}
 E.~Knill, R.~Laflamme, and G.~J.~Milburn,
 Nature {\bf 409}, 46 (2001).

 \bibitem{Giovannetti2004}
 V.~Giovannetti, S.~Lloyd, and L.~Maccone, Science {\bf 306}, 1330 (2004).

 \bibitem{Varcoe2000}
 B.~T.~H.~Varcoe, S.~Brattke, M.~Weidinger, and H.~Walther,
 Nature {\bf 403}, 743 (2000).

 \bibitem{Wallraff2004}
 A.~Wallraff, D.~I.~Schuster, A.~Blais, L.~Frunzio, R.-S.~Huang,
 J.~Majer, S.~Kumar, S.~M.~Girvin, and R.~J.~Schoelkopf,
 Nature {\bf 431}, 162 (2004).

 \bibitem{Delft2004}
 I.~Chiorescu, P.~Bertet, K.~Semba, Y.~Nakamura, C.~S.~P.~M.~Harmans,
 and J.~E.~Mooij, Nature {\bf 431}, 159 (2004).

 \bibitem{Schuster2005}
 D.~I.~Schuster, A.~Wallraff, A.~Blais, L.~Frunzio, R.-S.~Huang,
 J.~Majer, S.~M.~Girvin, and R.~J.~Schoelkopf,
 Phys.~Rev.~Lett.~{\bf 94}, 123602 (2005).

 \bibitem{Schuster2007}
 D.~I.~Schuster, A.~A.~Houck, J.~A.~Schreier, A.~Wallraff,
 J.~M.~Gambetta, A.~Blais, L.~Frunzio, J.~Majer, B.~Johnson,
 M.~H.~Devoret, S.~M.~Girvin, and R.~J.~Schoelkopf,
 Nature {\bf 445}, 515 (2007).

 \bibitem{Blais2004}
 A.~Blais, R.-S.~Huang, A.~Wallraff, S.~M.~Girvin, and R.~J.~Schoelkopf,
 Phys.~Rev.~A {\bf 69}, 062320 (2004).


\bibitem{Jena2003}
E.~Il'ichev, N.~Oukhanski, A.~Izmalkov, Th.~Wagner, M.~Grajcar,
H.-G.~Wagner, A.~Yu.~Smirnov, A.~M, van den Brink, M.~H.~S.~Amin,
and A.~M.~Zagoskin, Phys.~Rev.~Lett.~{\bf 91}, 097906 (2003).


\bibitem{Hauss2007}
J.~Hauss, A.~Fedorov, C.~Hutter, A.~Shnirman, and G.~Sch\"on,
Phys.~Rev.~Lett. {\bf 100}, 037003 (2008).

\bibitem{Astafiev2007}
D.~Astafiev, K.~Inomata, A.~O.~Niskanen, T.~Yamamoto,
Yu.~A.~Pashkin, Y.~Nakamura, and J.~S.~Tsai, Nature {\bf 449}, 588
(2007).

\bibitem{Rodriguess2005}
D.~A.~Rodrigues, and A.~D.~Armour, New J.~Phys.~{\bf 7}, 251
(2005).

\bibitem{Blencowe2005}
M.~P.~Blencowe, J.~Imbers, and A.~D.~Armour, New J.~Phys.~{\bf 7},
236 (2005).

\bibitem{Clerk2005}
A.~A.~Clerk, and S.~Bennet, New J.~Phys, {\bf 7} 238 (2005).


\bibitem{Rodrigues2007}
D.~A.~Rodrigues, J.~Imbers, and A.~D.~Armour,
Phys.~Rev.~Lett.~{\bf 98}, 067204 (2007).


\bibitem{Fulton1989}
T.~A.~Fulton \emph{et.~al.}, Phys.~Rev.~Lett.~{\bf 63}, 1307 (1989).

\bibitem{Maassen1991}
A.~Maassen van den Brink, G.~Sch\"on, and L.J.~Geerligs,
Phys.~Rev.~Lett.~{\bf 67}, 3030 (1991); A.~Maassen van den Brink,
A.A.~Odintsov, P.A.~Bobbert, and G.~Sch\"on, Z.~Physik B -
Condensed Matter {\bf 85}, 459 (1991).

\bibitem{Almog2007}
R. Almog, S. Zaitsev, O. Shtempluck, and E. Bucks, Phys. Rev.
Lett. {\bf 98}, 078103 (2007).

 \bibitem{Makhlin2001}
 Y.~Makhlin, G.~Sch\"on, and A.~Shnirman,
 Rev.~Mod.~Phys.~{\bf 73}, 357 (2001).

 \bibitem{Choi2003}
 M.-S.~Choi, F.~Plastina, and R.~Fazio, Phys.~Rev.~B.~{\bf 67},
 045105 (2003).

 \bibitem{Naik2006}
 A.~Naik, O.~Buu, M.~D.~LaHaye, A.~D.~Armour,
 A.~A.~Clerk, M.~P.~Blencowe, and K.~C.~Schwab,
 Nature {\bf 443}, 193 (2006).

\end{thebibliography}
\end{document}